\documentclass{svmult}

\usepackage{makeidx}         
\usepackage{graphicx}        
\usepackage{multicol}        
\usepackage[bottom]{footmisc}

\begin{document}

\title*{Systemic Risk: Fire-Walling Financial Systems Using Network-Based 
Approaches}
\titlerunning{Systemic risk in financial networks}
\author{V. Sasidevan \inst{1}\and
Nils Bertschinger\inst{2}}
\institute{Department of Physics,  Cochin University of Science and Technology, 
Cochin - 682022  India.
\texttt{sasidevan@cusat.ac.in, sasidevan@gmail.com}
\and {Frankfurt Institute for Advanced Studies, Frankfurt am Main, 
Germany.\\
Department of Computer Science, Goethe University, Frankfurt am Main, Germany.}
\texttt{bertschinger@fias.uni-frankfurt.de}}
%
%
\maketitle
\begin{abstract}
 The latest financial crisis has painfully revealed the dangers 
arising 
from a globally interconnected financial system. Conventional approaches based
on the notion of the existence of equilibrium and those which rely on
statistical forecasting have seen to be inadequate to describe financial
systems in any reasonable way. A more natural approach is to treat fi-
nancial systems as complex networks of claims and obligations between
various financial institutions present in an economy. The generic frame-
work of complex networks has been successfully applied across several dis-
ciplines, e.g., explaining cascading failures in power transmission systems
and epidemic spreading. Here we review various network models address-
ing financial contagion via direct inter-bank contracts and indirectly via
overlapping portfolios of financial institutions. In particular, we discuss
the implications of the \textquotedblleft 
robust-yet-fragile\textquotedblright\, nature of financial networks
for cost-effective regulation of systemic risk.
\end{abstract}

\section{Introduction}
\label{sec:1}
Over the past few decades, the generic framework of complex networks has 
been used to address problems in a variety of scientific disciplines. 
The study of evolutionary dynamics of populations \cite{lieberman2005}, 
spreading of diseases \cite{newman2002}, spreading of failures in power 
transmission systems \cite{dobson2007}, dynamics of opinion formation 
\cite{castellano2009}, working of genetic regulation \cite{hasty2001}; all 
heavily rely upon viewing and understanding the underlying systems 
as complex networks which represent the interaction between their various 
constituent elements. It is now well-known that the emergent or collective 
behavior in these systems are greatly influenced by the topology of the 
underlying network of connections together with the particular kind of dynamic 
processes which run on it \cite{strogatz2001, amaral2004, boccaletti2006}. 

One of the latest entry into the class problems to be analyzed using methods 
based on the 
theory of complex networks is the issue of spreading of failures in financial/economic 
systems. Often dubbed as systemic risk, this refers to the possibility of spread 
of initial distress at a single or a few financial institutions such as banks 
to a significant fraction of a financial network.  Though earlier studies of 
such contagion exist, it is safe to say that it is only in the 
aftermath of the financial crisis of 2007-09 that viewing systems 
such as interbank markets from the standpoint of the theory of complex networks 
became mainstream \cite{schweitzer2009,haldane2015}.

More and more direct and indirect connections are being 
made everyday between financial institutions as the world increasingly becomes 
an economically globalized one. This together with modern day financial 
products such as CDS (credit default swaps) or CDO (collateralized debt obligations)
make today's financial systems a very 
complex one and hard to analyze. Conventional approaches based on the notion of 
the existence of equilibrium and those which rely on statistical forecasting 
have seen to be inadequate to describe financial systems in any 
reasonable way \cite{farmer2009}. Conventional theory often makes unrealistic 
assumptions and is capable of analyzing only simple situations. Most 
importantly, it cannot handle non-linearities and feedback 
mechanisms ubiquitous in financial systems \cite{catanzaro2013}. Inadequacy of 
the 
conventional approaches together with the spread of concepts from complex 
network theory to the economic/financial community lead to the conclusion that a 
more natural way to describe financial systems is to treat them as complex 
networks of claims and obligations between various financial institutions 
present 
in an economy. It is to be noted that a significant push towards this 
direction was provided by people working on ecological and epidemiological 
systems (e.g., Robert May \cite{may2008}). The result is that there has been an 
exponential 
increase in the studies of financial networks employing ideas and tools from the  
complex networks literature and the terminology of latter nowadays finds 
commonplace in the discussions related to contagion in financial systems.

In the simplest form, the nodes of a financial network are institutions involved 
in financial intermediation (called banks in short hereafter) and the links
between the nodes represent lending/borrowing relationships. The links are 
therefore directed (usually represented as directed from borrower to 
lender indicating the anticipated direction of cash flow after the link has 
been established) and weight of a link 
represents the amount of money involved in the transaction between the nodes 
connected by that link. The banks may have other kinds of direct and indirect 
connections between them which are discussed in section 2. 
Given such a financial network, the term systemic risk refers to the 
possibility of an initial 
financial distress at a single or a few banks (say their defaulting due to a 
loss of value of their asset holdings) 
to spread through the network causing the default of a significant number of other banks in the network. 
Such a  system wide collapse has serious repercussions for the financial system
and global economy in general and could affect the lives of millions in a negative way as painfully revealed in the 
crisis of 2007-09. 

We can easily see the parallels of the problem of the collapse of a financial 
network with that of an ecological network or the spread of an 
infectious disease. Although all these problems deal with the transmission of a 
\textquoteleft failure\textquoteright\, 
at a node to its neighbors in a general
sense, there are also marked differences between them \cite{may2011}.  A feature 
which distinguishes financial networks from others is that 
unlike the failure of a node in an ecological network or an epidemiological 
network, the failure of a node in a financial 
network depends upon the  \textquoteleft internal structure\textquoteright\, of 
that node (i.e., balance sheet structure of the bank) and the 
\textquoteleft cumulative health\textquoteright\, of its neighbors in the 
network. So unlike simple stochastic 
rules used to model for e.g., the spread of an infectious disease (as in SIR 
model), here we 
need to consider the internal balance sheet dynamics of a node in detail.  

Our aim in this review is to give a broad overview of the various developments 
in 
the field of systemic risk of financial networks and discuss some future 
directions. In this respect, we will mainly focus on the 
literature that concentrates on the use of tools from the study of complex 
networks.  There is an extensive literature on banking 
systems based on more traditional point 
of views like the ones using equilibrium models.
We will occasionally mention other strands of research but will refrain from 
discussing them in any detail. 

This review is organized as follows. In section \ref{sec:2}, we give an 
account of the structure of a financial network and a detailed description 
of what constitutes its nodes and links. We describe the 
kind of questions one would like to answer about such networks. In section 
\ref{sec:3}, we review 
various past efforts in modeling contagion via interbank credit networks and 
in section \ref{sec:4}, we review contagion via fire-sale of commonly held 
assets by banks. We discuss some of our findings related to asset contagion. In section \ref{sec:6} 
we conclude and discuss some future possibilities. 

\section{Nodes and links of a financial network}
\label{sec:2}
\begin{figure}[t]
 \centering
 \includegraphics[scale = 0.8]{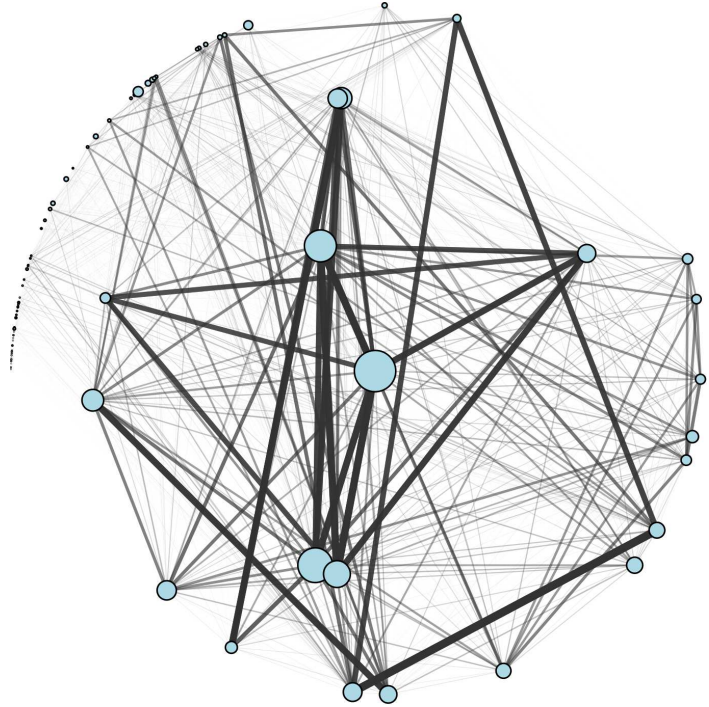}
 \caption{A graphical representation of the banking network of major European 
banks created using data from the European Banking Authority (EBA). The size of 
the nodes represent the relative size of the balance sheets of the banks, and 
the thickness of the links represents the value of common asset exposures.}
 \label{eu_network}      
 \end{figure}
\begin{figure}[t]
 \centering
 \includegraphics[width=\textwidth]{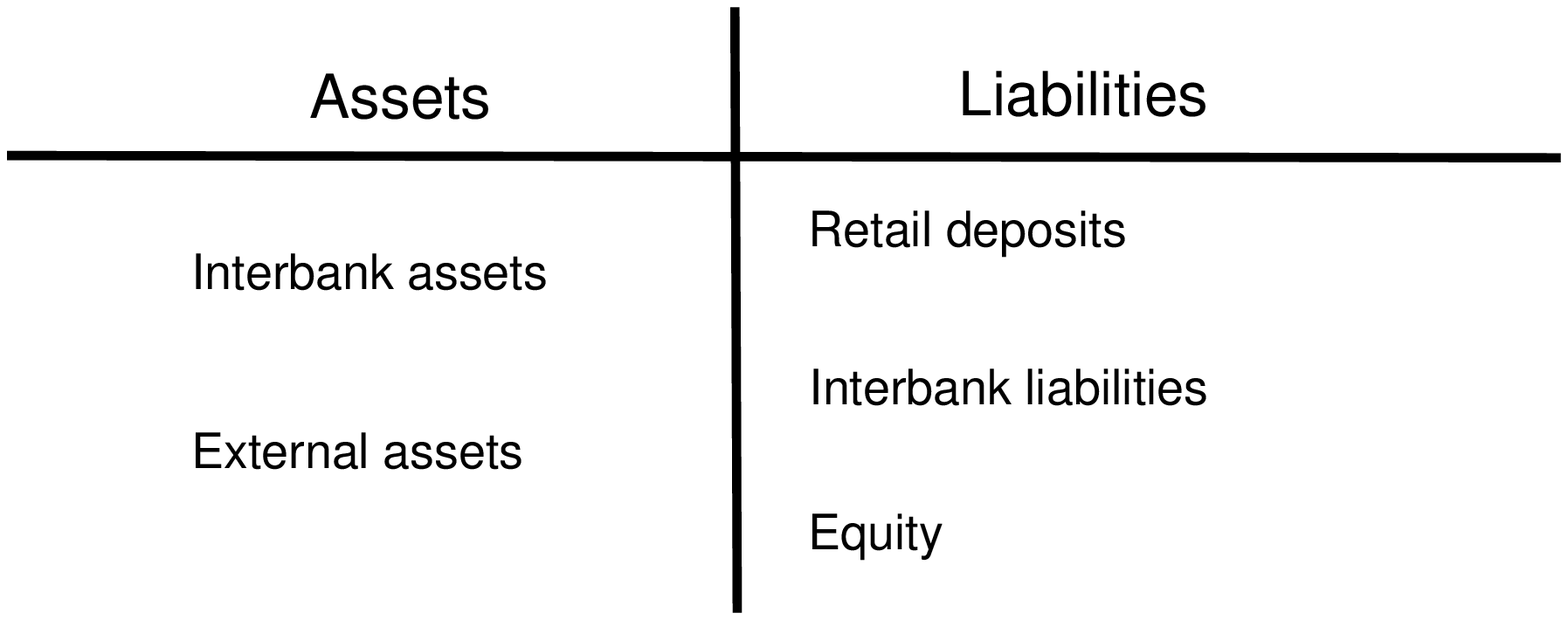}
 \caption{A simplified schematic of a balance sheet of a bank indicating a 
minimal division of its assets and liabilities into various classes. Interbank 
assets represent the money that the bank lent to other banks and external 
assets represents the likes of money lent to industry. Interbank liabilities 
represent the money that is owed to other banks. Finer sub devisions of assets 
and liabilities can be considered depending upon the specific contexts (see 
text). The difference between the total assets and liabilities gives its 
equity. The bank is  insolvent if its net liabilities exceed the value of its 
assets.}
 \label{balance_sheet}      
 \end{figure}

As mentioned in the introduction, the nodes of a financial network are banks, 
and the links represent various credit relationships. A sample financial 
network based on data from the European Banking Authority (EBA) is given in 
Fig.~\ref{eu_network}.  
To study the propagation of \textquoteleft 
failures\textquoteright\, or \textquoteleft defaults\textquoteright\, in a 
financial network, we may represent a 
bank by its balance sheet structure which lists its various assets 
and liabilities at any time. A schematic of a balance sheet is 
shown in Fig.~\ref{balance_sheet}. Broadly, we may classify the assets of a 
bank 
into two types; 
namely interbank assets (the money that the bank lent to other banks) and 
external assets (investments like the money that the bank lent to other parties 
such as business firms often receiving a collateral, deposits with a 
central bank, etc). The external assets can be 
further classified into liquid assets (assets for which there is a market and 
hence can be easily converted into cash. Examples include cash, central bank 
reserves, high-quality government bonds, etc.)) and illiquid ones (like secured 
or unsecured mortgages 
which cannot be used to raise cash quickly). More 
detailed classification of assets is possible by further considering 
divisions such as \textquoteleft Collateral assets\textquoteright\, (assets 
which may be used as 
collateral in repo transactions) and \textquoteleft Reverse repo 
assets\textquoteright\, (i.e., collateralized 
lending), etc. \cite{gai2011}. Considering such divisions is vital as the 
percentage of assets in various classes decides whether a bank in need of 
immediate cash for its promised payments can raise it in time or 
not. In general, the size of the balance sheet, i.e., the total value of assets 
(or 
equivalently liabilities) varies across banks. It is often 
found that the size distribution is heterogeneous having a heavy tail 
\cite{raddant2016}. 

The liability side of the balance sheet usually consists of interbank 
liabilities (money that is owed to other banks) and retail deposits. Finer 
divisions such as \textquoteleft repo liabilities\textquoteright\, (i.e., 
borrowing secured with collateral) may 
also be considered \cite{gai2011}. The difference between the total assets and 
liabilities of a bank gives its equity, also commonly called net worth or capital. The equity is written on 
the liability side of the balance sheet indicating that it is money owed to promoters and 
shareholders of the bank.  

There are two important considerations regarding the \textquoteleft 
health\textquoteright\, of a node.
\begin{itemize}
 \item \textbf{Insolvency:} A bank is considered insolvent if it has a negative 
net worth. The net worth may become 
negative, say for e.g., due to a drop in the prices of its assets (in a 
mark-to-market framework, the value of assets are updated according to its 
current 
market price) or due to failed investments. We may say that the bank cannot 
operate anymore as it cannot honor its liabilities and should close down with 
appropriate insolvency proceedings, though in reality, a number of things may 
happen like possible government intervention, etc.
\item \textbf{Illiquidity:} A bank is illiquid if its short term liabilities 
exceed liquid assets and 
hence it cannot honor the former. 
\end{itemize}

In both cases, the bank is in trouble, and this will affect other 
banks to which it owes money. Also, the assets of the bank that is in trouble 
will face devaluation, thus affecting other banks holding the same class of 
assets on their balance sheets. Due to such feedbacks, initial 
distress can spread to other banks in the network and may affect a large part of 
it. 

\begin{figure}[t]
 \centering
 \includegraphics[width=\textwidth]{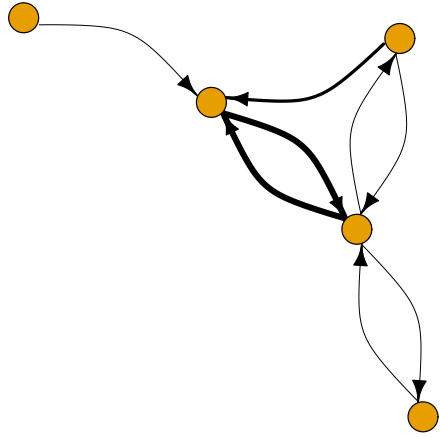}
 \caption{A directed and weighted network consisting of 5 banks. The 
thickness of the links represent the value of credit exposures.}
 \label{directed_network}      
 \end{figure}

The links between nodes represent lending/borrowing relations between banks. 
They are directed and weighted, the weight of a link representing the value 
owed (See Fig.~\ref{directed_network}). Note that banks may interact with 
others 
not only through the above 
lending/borrowing relationship. Their balance sheet may be affected by the fact 
that they are exposed to common assets. A sell off or even a negative news 
about  a bank might cause depreciation in the value of the assets it 
holds and other banks which have exposure to these assets will be affected. 
The sell off of an asset by a distressed bank will cause excess supply of that 
asset and if the market depth of that asset is shallow, will lower its value 
significantly. Such fire sale losses can quickly spread among the banks in the 
network. These two mechanisms of interbank lending and fire sale of 
assets form two important channels for spreading of 
crisis in a 
financial network. There are other channels possible and which ones are more 
relevant in reality is debatable. Upper \cite{upper2011} for e.g., argues that 
there is not much evidence for interbank channels propagating distress. In any 
case, these are possible channels of contagion under conceivable circumstances, 
and hence it is important to understand them from a regulatory and individual 
point of view.

Being able to understand the \textquoteleft behavior\textquoteright\, of a 
bank is crucial in the study of systemic risk. This is because the way 
a bank organizes its balance sheet and how they form connections will have a 
significant role in deciding whether or not an initial default will propagate in 
the network 
causing significant damage. An important factor which determines the 
actions of a bank is various  regulatory constraints set by a central authority 
(like a central bank). For e.g., there could be a capital ratio constraint set 
by the authority stipulating that the activities of the bank should obey the 
constraint that its capital or net worth should be a minimum percentage of its 
illiquid assets (8\% is now a typical figure under Basel III). Also the bank itself may have 
policies to keep its balance sheet structure in a specific way (For e.g., 
maintaining a constant leverage ratio \cite{adrian2010}).

\section{Contagion via interbank linkages}
\label{sec:3}
Lending and borrowing relationships with other banks serve the 
purpose of meeting the liquidity needs of a bank and profiting from the lending 
of idle liquid money. While lending, there is, of course, the risk of default. 
Presumably, forming more connections or 
dividing total exposure across several banks  will reduce such risk. 
On the 
other hand, a bank with more number of connection will also have a higher 
probability of being hit by a defaulting counterparty. The interplay between 
these two effects determines whether the default of a single or a small set of 
banks will spread through the system.

\subsection{Contagion via default spreading}

One of the earlier works which explicitly incorporated the network aspects of 
interbank lending/borrowing is by Allen \& Gale \cite{allen2000}. The interbank 
network they considered is rather simple compared to any real-world 
banking network and involved only four banks. Two different 
connectivity structures of the network are considered, and they showed that a 
complete 
network in which each bank is connected to every other bank readily absorbs any 
initial shock. However, an incomplete network structure makes the system prone 
to the spread of an initial default.

A pioneering work which considered the effects of varying the 
important features of interbank networks is by Nier et al. \cite{nier2007}. The 
network 
structures considered were Erdos-Renyi (ER) type and a tiered one. In the 
latter, 
a few large banks are connected to a large number of small banks where the 
latter are mostly connected to the large ones. The effect 
of level of capitalization of the banks, degree of the ER network, size of the 
interbank exposures, etc. on the number of defaults is determined using 
simulations. It is 
found that at smaller degrees of the network, a small increase in the 
degree of the network causes an 
increase in the number of defaults, but at relatively higher degrees, the 
effect of an initial default is contained. Predictably, higher capital levels 
found to reduce the risk of contagion, but the effect is found to be 
non-linear. The paper also briefly considers the \textquoteleft fire 
sale\textquoteright\, channel in which a 
distressed bank sells off its asset causing a depreciation in the market value 
of the assets. 

\begin{figure}[t]
 \centering
 \includegraphics[width=\textwidth]{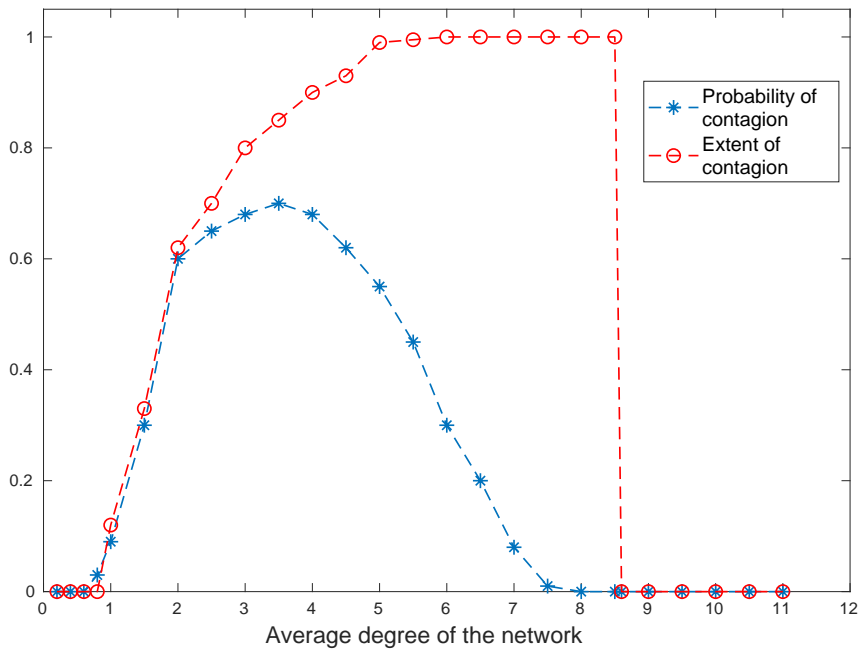}
 \caption{A schematic representation of the dependence of probability of 
contagion and extend of contagion on the average degree of an ER network.}
 \label{contagion}      
 \end{figure}

A landmark work is that of Gai \& Kapadia \cite{gai2010} in which they 
formalized the above setting by Nier et al. \cite{nier2007} and distinguished 
probability of contagion from its extent (A comparison between these 
models is given in \cite{may2010}). The former denotes the probability of a 
systemic event 
conditional on an initial shock while the latter denotes the percentage of 
banks affected conditional on a systemic event. One should note that what 
constitutes a 
systemic event is somewhat subjective. In this paper and several later works, 
a 
contagion is considered as systemic if more than 5\% of the banks default. This 
might seem like a very low number, but reflects the fact that even the default 
of 5\% of banks is a severe event in an economy affecting a large number of 
people and potentially wiping out millions of dollars. Their theoretical 
analysis closely follows the study of cluster size distribution in a complex 
network using generating functional formalism (see for e.g. \cite{newman2010}). 
The key 
finding was that the phenomena of contagion in a financial network have a 
\textquoteleft robust-yet-fragile\textquoteright\, nature. i.e., even if the 
probability of contagion is very 
low, whenever a contagion happens, it wipes out most of the system. The typical 
dependency of probability and extent of contagion on the average degree of the 
ER network is as shown in Fig.~\ref{contagion}. Note that when the average 
degree is 
below 1, there is no possibility of contagion as the network is a disconnected 
one and there is no large connected component present. After that, the 
probability of contagion increases with an increase in average degree, reaches a 
maximum value and then decreases. This clearly shows the interplay of the two 
effects of risk sharing and default spreading mentioned earlier. Higher degrees 
diffuses the initial impact among many neighbors thus reducing the possibility 
of contagion. However, once a contagion starts to spread, higher degrees can 
act 
as an ideal setting for the propagation of defaults since many counter-parties 
of a 
bank will be simultaneously affected. In a nutshell, this has the effect that 
the banking network may respond in entirely different ways to similar shocks at 
different nodes in the network.

The assumption made in \cite{gai2010} regarding the network structure as being 
ER type and that all banks are of the same size is not in line with empirical 
evidence about real-world banking networks which are found to have heavy-tailed 
distribution for both connectivity and size of banks \cite{huser2015}. Caccioli 
et al. \cite{caccioli2012} extended the Gai \& Kapadia model to incorporate 
these 
effects. It was found that a heterogeneous degree distribution makes the network 
more resilient towards random defaults but more vulnerable to targeted attacks 
on 
well-connected nodes. It was also found that a power-law distribution for the 
size of the banks makes the contagion events more frequent. The effects of 
policies 
such as increased capital requirements are also considered which are seen to 
improve the resilience of the system. 
A recent work discussing the role of assumptions made 
about the connectivity and size of banks is by Raddant \cite{raddant2016} in 
which features of real-world empirical networks were considered in some 
detail.

\subsection{Contagion via valuation adjustments}

The seminal work of Eisenberg \& Noe \cite{eisenberg2001} provides an alternative viewpoint
on default contagion, namely as
a problem of contract valuation. While each bank has promised the repayment of
a nominal amount, it might be unable to hold up to its promise when the
payment is due. As each repaid credit -- in full or in part -- is an
income to the lending counterparty, Eisenberg \& Noe proposed that the
payments of all banks have to be considered simultaneously and solved
in a self-consistent fashion. The resulting {\em clearing vector} can
then be regarded as the value of all interbank contracts. Thus,
instead of modeling the spread of defaults directly, the model
provides a self-consistent solution for the final state when all
payments are settled. This point of view readily connects with more
standard approaches for financial valuation and has since spurred a
range of theoretical \cite{glasserman2015} and computational research
\cite{upper2011}.

In particular, the model of Suzuki \cite{suzuki2002} is noteworthy in
this respect.  It not only generalizes the above model allowing for
cross-holdings of equity as well as interbank debt but also clearly
formulates it as an extension of the Merton model to multiple firms.
In his ground-breaking work, Merton \cite{merton1974} showed that
liabilities and equity of a firm could be considered and valued as a
short \textit{put} and long \textit{call} option on the firm's assets
respectively. Nowadays this insight forms the foundation of structural
credit risk modeling and has been developed into established
industry practice. Valuation in a network context is considerably more
complicated as all option values are interdependent and have to be solved
in a self-consistent fashion. Correspondingly, almost no analytic
results are known, and only recently, some extensions including
interbank liabilities of different seniorities \cite{fischer2014} or
maturities \cite{kusnetsov2019} have been proposed.

From the perspective of a single firm, an approximate valuation can be
obtained by taking into account the solvency risk of its
counterparties. Barucca et al. \cite{barucca2016} have shown that
several network valuation models, including the Eisenberg \& Noe model
and many of its extensions, can be unified in this way. In their framework,
each firm locally adjusts the value of interbank contracts according
to its expected recovery value. Different valuation models are then
obtained via different assumptions about when nominal contract values
are adjusted -- either after an actual default or before
an expected default. In the first case, the valuations obtained by
each firm agree with the values derived from self-consistent network
considerations whereas they only approximately reflect market values
of interbank contracts in the latter.

Within the framework of network valuation, financial contagion is naturally quantified in
marginal terms, i.e., as the impact of infinitesimal asset price shocks
\cite{ota2014}. Demange \cite{demange2016} proposed a
threat index considering a bank as systemically important to the
extent that the values of all other banks are sensitive to devaluation
of its assets. Taking this idea even further, Bertschinger \& Stobbe
\cite{bertschinger2018} have shown how to compute network Greeks,
i.e., sensitivities with respect to several risk factors of interest
besides asset prices. In principle, this allows to quantify the
systemic impact of changes in interest rate, volatilities or asset
correlations and naturally extends standard risk management
practices. Indeed, the Greeks of single asset options are routinely
used in order to asses and hedge the risk of trading portfolios.

\subsection{Strategic reactions and network structure}

An assumption that is made in the studies mentioned above is that the banks 
remain passive in the face of spreading of defaults. While there is some merit 
to this argument, it is doubtful that banks remain  passive when aware of a 
possible default by a counterparty or news of trouble elsewhere in the financial 
system. There has been only minimal effort to model the effect of the 
response of banks on financial contagion. Anand et al. \cite{anand2012}  
incorporate a strategic decision-making component to the network which 
makes rewiring of connections a possibility. They model liquidity hoarding 
behavior of banks in which a bank refuses to roll over an interbank loan in fear 
of default of the counterparty. The outcome of a coordination game among the 
neighbors of a bank decides whether they roll over or withdraw loans made to 
that bank. Another work is by Arinaminpathy et al. \cite{arinaminpathy2012} in 
which confidence effects are included in the response of banks. Among other 
defining features of the model, it is assumed that long term loans can be made 
short term and interbank loans may be withdrawn.       

A natural question is what kind of network structure will increase the stability 
of a financial system. The question has been a much debated one regarding all 
complex systems from the time of the influential article by Herbert Simon 
\cite{simon1962}. Stability of many benchmark network structures of 
financial systems has been studied by Battiston et al. \cite{battiston2013} 
with 
the conclusion that network topology plays a prominent role only when the 
liquidity is low, and there is no single optimal connection structure.

\section{Contagion via fire sale of assets}
\label{sec:4}
As already mentioned, apart from the direct interbank claims and obligations, 
another way in which the balance sheets of banks influence each other is 
by carrying the same or similar asset classes. Portfolio overlaps thus produce 
another 
channel for initial defaults to spread, and evidence suggests that this one is 
more relevant than the direct channel
in real-world systems. Though several earlier studies like that by Gai 
\& Kapadia did consider minimally the effect of fire-sale of a single 
asset held by all 
banks, it is in more recent literature that this channel of 
contagion is analyzed more thoroughly. A general message here is that too much 
diversification 
may not be desirable for the health of the system even though it might appear 
as a 
wise strategy from an individual bank's point of view. In fact, prior to the 
2007-09 crisis, the prevailing view was that portfolio diversification by 
individual banks will make them safer and this will automatically translate 
into 
improved safety of the overall financial system. However, this ignored the 
crucial fact that diversification by firms will also create strong correlations 
across balance sheets thus rendering the whole system susceptible to price 
changes of commonly held assets. 

\begin{figure}[t]
 \centering
 \includegraphics[scale = .5]{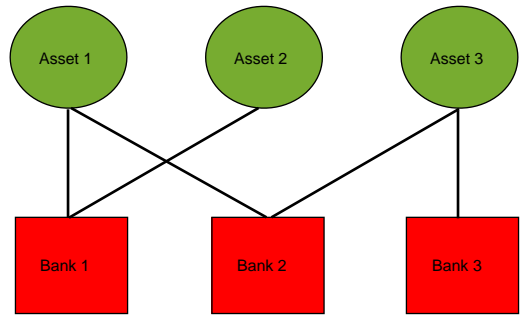}
 \caption{A schematic representation of a bipartite network of banks and 
assets. If Bank 1 liquidates its assets in a fire sale, it will depreciate the 
prizes of Asset 1 and Asset 2. This will affect Bank 2 as it has exposure to 
Asset 1. In case Bank 2 also forced to liquidate its assets, Bank 3 will be 
affected as it has exposure to Asset 3. Thus a contagion can spread due to 
correlated balance sheet structures }
 \label{banks_assets}      
 \end{figure}

\subsection{Contagion via overlapping portfolios}

An early work which 
focuses on the network aspects of portfolio overlaps is 
by Caccioli et al.  \cite{caccioli2014} who considered a bipartite network of 
banks and 
assets. In this network, banks have only connections to the assets they hold in 
their portfolio (see Fig.~\ref{banks_assets}). Two kinds of initial shocks are 
considered namely 1) 
Devaluation of 
the price of a randomly chosen asset to zero and 2) Making a randomly selected 
bank default.
The initial shock then causes defaulting banks to sell the assets in their 
portfolio en masse thereby depreciating their prizes. A specific functional 
form 
for the decrease in the price of an asset is assumed which in general could 
depend upon 
market depth and the  volume of assets on sale. The depreciating prizes could 
cause further 
defaults thereby initiating a cascade. The effects of leverage 
(asset to equity ratio), market crowding (number of banks per assets), 
diversification 
(as measured by the average degree of banks) and market impact (price change of 
an asset as a function of its volume) on the cascading process is studied. A 
major finding is that there is a critical leverage below which financial 
networks are immune to systemic events, but above which there is a finite 
probability for them. A theory based on generalized branching process is also 
developed. A point to note here is that the probability and 
extent of contagion due to fire sales behaves more or less in a similar way to 
that 
in the case of direct counter-party contagion as depicted in 
Fig.~\ref{contagion}.

Compared to interbank lending network, the asset contagion network can be more 
reliably constructed from the data available with regulatory authorities. Unlike 
interbank lending relations, the structure of the balance sheet of banks are 
partially
publicly available and can be used to construct empirical banks-assets 
networks. 
Huang et al. \cite{huang2013} ran an asset contagion model on US 
Commercial banks asset network. The latter is constructed using data  from US 
Commercial Banks Balance Sheet Data (CBBSD) which classifies assets into 
thirteen different types. By running simulations on the empirical network, 
they could 
reproduce a large portion of actually failed banks during the 2007-09 crisis. 
Similar balance sheet structure of major European Banks is made available 
by the European Banking Authority (EBA) and has been used in several
studies.  

Banwo et al. \cite{banwo2016} considered the effects of power-law 
distributions for the degree of the network as well as for balance sheet size. 
They found that the power-law degree distribution makes the system less 
resilient towards the initial default of a randomly chose bank. Targeted 
initial defaults  of biggest or less connected banks increase the probability 
of a default cascade. A power-law distribution for the assets, on the other 
hand, increases the resilience of the network with respect to random initial 
defaults but not with respect to targeted defaults. Caccioli et al. 
\cite{caccioli2015} studied the effect of both interbank and 
asset contagion channels working together. It is found that systemic 
effects 
are amplified when 
both channels are present.

\subsection{Fire-walling financial networks}
\label{sec:5}

By now it should be clear that a crucial question in these contexts is to 
determine which of the banks are 
most important as far as the stability of the network is concerned.  
Identifying systemic institutions is crucial in developing strategies 
and policies to protect the financial system from contagion. A 
simple answer to this is the ``too-big-to-fail'' paradigm in which it is proposed 
that the largest banks are the most important ones (which are also presumably 
well connected). A complimenting one to this is the 
``too-central-to-fail'' paradigm introduced by Battiston et al. 
\cite{battiston2012} 
where they 
introduced a centrality measure called Debt rank. Another centrality measure 
is the Systemicness of a bank introduced by Greenwood \cite{greenwood2015} 
which 
is proportional to the product of the size of a bank, its leverage, and 
connectedness. Information regarding these variables may not be readily 
obtainable, and Gangi \cite{gangi2015} discusses how to obtain Systemicness 
based on only partial data about the network. A recent addition to the list of 
various centrality measures is the one proposed by Cont \& Schaanning 
\cite{cont2016} which is based on liquidity weighted portfolio overlaps.
Also, the threat index by Demange \cite{demange2016} and the network $\Delta$ 
proposed by Bertschinger \& Stobbe \cite{bertschinger2018} are noteworthy in 
this context.

\subsubsection{Cost of preventing contagion}

\begin{figure}[t]
 \centering
\includegraphics[width=0.8\textwidth]{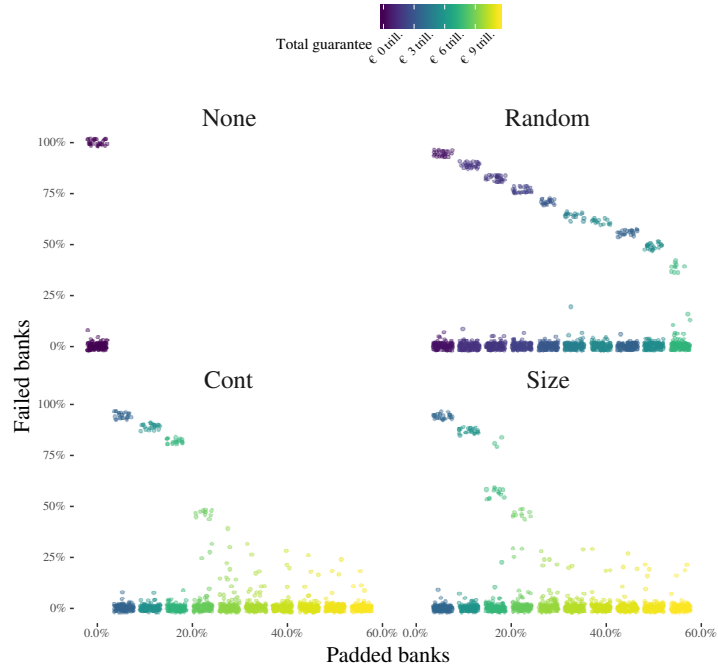}
 \caption{Effectiveness of different bail-out strategies to
  prevent contagion from the fire sale of assets. Bail-out guaranties are
 provided to a fraction of all banks (padded banks) ranked either
   randomly (top-right) or according to two different centrality measures. The 
two centrality measures considered are the size of banks (bottom-right) and 
the one by Cont \& Schaanning based on liquidity weighted portfolio overlaps of 
banks (bottom-left) \cite{cont2016}. The case with no bail-out guarantees is 
also shown (top-left).}
 \label{fig:bailout}      
 \end{figure}

The proposed centrality measures should not just provide a better understanding 
of the systemic importance of financial institutions, but also should play a 
key role in designing cost-effective countermeasures to curtail the financial 
crisis. Here, we present a simulation study illustrating this idea. In 
particular, we study the model of Cont \& Schaanning using portfolio holding 
data from the EBA (see \cite{cont2016} for additional details). The data set
contains information about investments of the 90 largest European
banks in about 140 asset classes, most finely disaggregated across
European government bonds. We consider a shock scenario
where a single asset is devalued by 30\% and simulate ensuing banking defaults
resulting from multiple rounds of fire-sale contagion. We
investigate two ways of fire-walling the system against financial
contagion: First via direct bail-out guaranties and second via asset
price guaranties.

Fig.~\ref{fig:bailout} shows the effectiveness of different bail-out
strategies. Here, each dot corresponds to a shock scenario and is
slightly jittered to reduce over-plotting. The all-or-nothing
character of contagion is clearly visible with either almost no or all
banks defaulting. Furthermore, providing bail-out guarantees to the
most central banks first, i.e., 10\% with the highest centrality, is
more effective in preventing contagion than rescuing banks at
random. Still, to fully avoid a systemic crisis guarantees on
more than 5 trillion euros have to be granted towards more than 30
major banks (30\% of 90 banks).

\begin{figure}[t]
 \centering
 \includegraphics[width=0.8\textwidth]{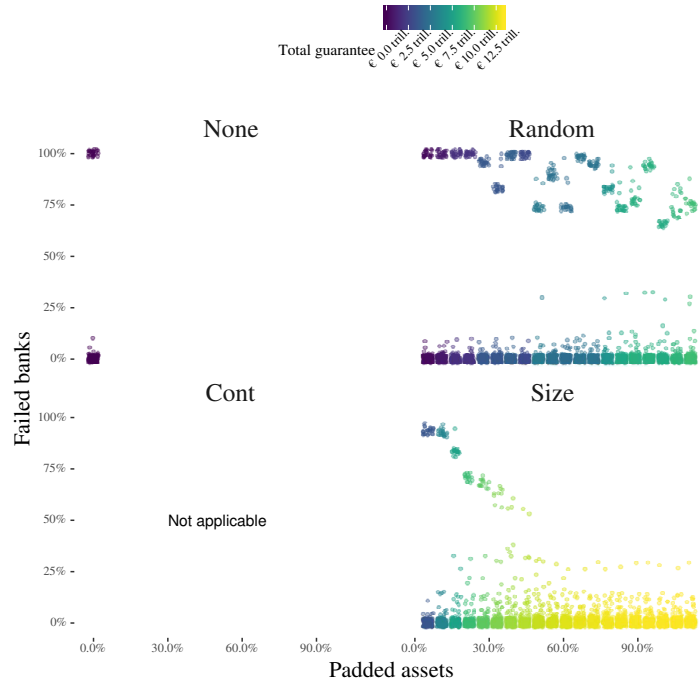}
 \caption{Effectiveness of different buy-out strategies to
   prevent contagion from the fire sale of assets. Buy-out guaranties are
   provided to stabilize the prices of a fraction of all assets
   (padded assets) ranked either randomly (top-right) or according to the 
volume of assets (bottom-right). The centrality measure proposed by Cont \& 
Schaanning based on liquidity weighted portfolio overlaps of banks
\cite{cont2016} is not applicable here. The case with no buy-out guaranties is 
also shown (top-left).}
 \label{fig:buyout}      
 \end{figure}

Fig.~\ref{fig:buyout} shows the corresponding results of different
buy-out strategies. In this case, instead of rescuing
individual banks, prices of certain assets are guaranteed. Thus, even
in case of a fire-sale, no price impact occurs on the guaranteed assets 
preventing any contagion via them. Again, providing buy-out guarantees 
in the order of decreasing systemic importance is found to be more effective 
than randomly buying them out. Still, at comparable sizes, buy-out guaranties
are found to be less effective in preventing contagion than direct bank 
bail-outs. Note that the nominal size of the guarantee -- as stated here and
often quoted by the media during the financial crisis -- can be much
larger than the actual implementation cost \cite{sasi2019}
especially if contagion is prevented.

\section{Conclusion and Future directions}
\label{sec:6}
The rapidly increasing connectivity and complexity of the global financial 
system is a major concern in today's world as it allows problems originating in 
one part to be quickly transmitted to seemingly distant and 
uncorrelated parts. Thus it has become ever more critical to understand the 
collective behavior of financial systems so that we can prevent or minimize 
contagious events. Viewing financial systems as complex 
networks is a significant step in this direction. The importance of 
macroprudential regulation in which more consideration is given to the health 
of the entire system than of individual institutions has now been recognized and 
is increasingly becoming a part of regulatory action. 

In this review, we have 
tried to give a broad overview of various attempts to understand and model
financial networks. A major theme arising out of these studies is the 
\textquoteleft robust-yet-fragile\textquoteright\, character of financial 
networks where contagious events are rare, but when they occur, they are 
widespread. Probability of 
contagion may be reduced by various regulatory measures such as
requiring higher capital ratios and higher liquidity ratios. Identifying 
systemic institutions in a financial network  is very important, and various  
centrality measures have been proposed for that. Protecting systemic 
institutions is crucial for the health of the entire system, but an issue here 
is how to set policies  such that the institutions behave in a responsible way 
towards the health of the entire financial system. There is the problem of 
\textquoteleft moral hazard\textquoteright\, where a systemically important 
institution, with the full knowledge that it will be rescued in a crisis by 
taxpayers money, engages in reckless risk-taking behavior.  

Financial networks are ever 
evolving and have continuously been subjected to many internal and external 
influences like regulatory constraints and factors from the economy. The 
technology underlying the formation of financial networks itself could change in 
the coming years with the use of Artificial Intelligence and formalizing 
financial contracts \cite{brammertz2018}. Another area to watch out for 
is the increasing presence of cryptocurrency \cite{borri2019}. All these have 
the potential to change the global landscape of financial networks. Including 
the effects of these and other future possibilities is a major challenge going 
forward. As we mentioned earlier, including the behavior of the banks in the 
face of a potential crisis is essential to model banking systems realistically. 
An area in which the current models lack realism is in the exclusion of central 
banks from modeling as the presence and actions of such a bank can have a 
significant effect on systemic risk \cite{georg2013}. 

Banks and regulatory authorities often do not have a complete picture of 
various direct and indirect exposures of institutions in a financial 
network \cite{battiston2012a}. Many recent studies try to integrate theoretical 
models with the available data to come up with conclusions about the 
possibility of systemic events and strategies or policies to contain them. With 
the availability of more and more real-time data and realistic models of bank 
behavior, it is expected that effective real-time monitoring of the health of 
financial systems will be a reality in the near future.
Together with innovative taxation schemes, such as the one proposed by Poledna 
\& Thurner \cite{poledna2016},
this might pave the way towards an economic future without systemic financial 
crises.
\section*{Acknowledgement}
VS acknowledges support by University
Grants Commission-BSR Start-up Grant No:F.30-
415/2018(BSR). NB thanks Dr. h.c. Maucher for funding his position.

\printindex
\end{document}